\documentstyle[12pt,aasms4,psfig,epsf]{article}

\def \etal {et al.\thinspace}

\def \ne{$N_e$\thinspace }

\def \cm3{cm$^{-3}$\thinspace }
\def \s1{s$^{-1}$\thinspace }

\begin{document}

\title{Recombination spectra of Helium-Like Ions}
\author{M. A. Bautista and T. R. Kallman}
\affil{Laboratory for High Energy Astrophysics}
\affil{NASA Goddard Space Flight Center} 
\affil{Greenbelt, MD 20771, USA}

\begin{abstract}

We calculate the recombination spectra of the He-like ions He~I, C~V,
N~VI, O~VII, Ne~IX, Mg~XI, Si~XIII, S~XV, Ar~XVII, Ca~XIX, and Fe~XXV. 
We include the following physical processes:
radiative recombination, dielectronic recombination, three-body
recombination, electron impact ionization, and collisional excitation by electrons, protons, and 
$\alpha$-particles. The calculations also account for the effects of lowering of
the continuum at high densities and 
high density corrections to dielectronic recombination. 

From the populations of all levels 
in the recombined ions we construct models for He-like ions for fast computation of
their  spectra. Every model includes 29 bound levels up to n=5,
a pair of "superlevels" which accounts for radiative and collisional 
cascades from highly excited levels, 6 doubly excited levels that account
for the most important satellite lines, and a level that represents the 
hydrogenic recombining ion. The models are constructed in a 
way that allows for proper approach to LTE under appropriate conditions.
These models can simultaneously solve for the H/He-like ionization 
balance in photoionized or collisionally ionized plasmas and compute 
emission spectra including the combined effects of radiative and
dielectronic recombination, collisional excitation, photoionization from
excited levels, fluorescence,
and line trapping. The models can be used for any temperature 
between 100 and $10^9$K and electron densities of up to $10^{18}$ \cm3. 
The models can be easily used within spectral modeling codes or as 
stand-alone tools for spectral analysis. 

We present comparisons between the results of the present models and 
previous work. Significant differences are found between the present effective 
recombination rate coefficients to the $n=2$ and those of previous
estimates. Later, we study 
various emission line ratio diagnostics under collisional ionization 
and photoionized conditions.

\end{abstract}


\section{Introduction} 

Spectral lines from He-like ions are prominent throughout the 
electromagnetic spectrum. He~I is responsible for many lines in 
the optical and infrared spectra and they are commonly observed 
in nebular plasmas. Thus, the low temperature recombination spectrum (without 
dielectronic recombination) of
He~I has been extensively studied for over four decades (Mathis 
1957; Burgess \& Seaton 1960a,b; Pottasch 1961; Robbins 1968, 1970; 
Robbins \& Robinson 1971; Brocklehurst 1972; Bhatia \& Underhill 1987;
Almog \& Netzer 1989; Smits 1991, 1996; Hummer \& Storey 1998; 
Benjamin, Skillman \& Smits 1999). The tabulated results from these
calculations have then been extensively used in the analysis of observed
spectra. Alternatively, a few models for the HeI system 
have been constructed, based on the treatment by Cota (1987), for
implementation into plasma modeling codes (e.g. CLOUDY by Ferland 1986;
ION by Netzer 1987; XSTAR v.1 by Kallman \& Krolik 1995). 

The emission from He-like ions other than He~I 
is important in the X-ray spectra of astronomical and laboratory 
plasmas. Although the spectral lines from these ions that 
result from the n=2 levels 
have received much attention the treatment of recombination excitation
has alway been approximated, while no detailed calculation of 
these ions recombination spectra has been reported.
The theory for computing the line intensities for transitions from the
n=2 levels in coronal plasmas was developed by Gabriel \& Jordan (1969,
1970, 1972, 1973). This theory includes a treatment of dielectronic 
recombination contributions in addition to collisional excitation.
The next major advance in the subject was made by Mewe \& Schrijver 
(1978a,b,c) who introduced approximated treatments for dielectronic 
and radiative recombination, collisional excitation, and inner-shell 
ionization. All of this was based on poorly known atomic data which they
fit according to ionic nuclear charge along the whole isoelectronic
series. Pradhan (1985) studied again the emission from n=2 states paying
special attention to recombination contributions to the emission in
non-coronal plasmas. 
Thus, most currently available 
spectra modeling codes simply extrapolate effective recombination rates
for any He-like ion from the  He~I rates using
hydrogenic scaling rules. 

There are at least two other difficulties with available He-like emission 
models and
calculations. One is that all of them are restricted to densities less 
than $10^{14}$ \cm3. Yet, various kinds of astronomical X-ray sources
of interest could be much denser, for example low-mass X-ray binaries 
(Bautista \etal 1998) and accretion disks near extragalactic black holes
(Nayakshin, Kazanas, and Kallman 2000). The other difficulty is that none of the calculations and
models correctly approach LTE under proper conditions. This invalidates
those results in the presence of strong radiation fields and in nearly
optically thick plasmas. 

In the present paper we start by presenting the atomic data used for the
calculation. Then we described  the calculations of 
level population and recombination spectra. In Section 4, we explain how
the recombination-collisional-radiative models were constructed.
In Section 5  
we compare the present models and their results with previous work. Finally, our conclusions are
presented in Section 6.

\section{Atomic Data for He-like ions} 
 
In this section we describe the atomic data used in the present work.
It is important to note that all of the present calculations are carried
out in LS coupling since fine-structure recombination rate coefficients are unavailable.
The LS approximation is expected to be accurate as long as $\Delta n$=0
radiative transitions are small and
relativistic
effects in the photoionization cross sections and radiative transition
rates remain small.
Although the present LS calculations are expected to be accurate 
additional work is currently in progress to compute 
the atomic data necessary for fine structure recombination calculations.

\subsection{Energy levels and line wavelengths}

Term energies and line wavelengths are taken from the compilation of
the National Institute of Standards and Technology (Martin, Sugar, and
Musgrove 1999)
when available. Energies for other $l\le2$ terms are from the TOPbase. 
Energies for $l> 2$ terms not known experimentally 
are assumed as hydrogenic. 

\subsection{Photoionization cross sections and recombination rates}

We use the photoionization cross sections for n$\le10$ and $\l\le2$ of 
Fernley, Taylor, \& Seaton (1987) calculated as part of the Opacity Project 
(OP hereafter; Seaton 1987) and which are available from TOPbase 
(Cunto \etal 1993). In principle one could obtain total, radiative and
dielectronic,  recombination 
rate coefficients by means of the Milne relation using cross sections
that resolve all autoionizing resonances. There are, however, some 
limitations in the OP cross sections: (1) the cross sections are
tabulated with an energy mesh too coarse to properly resolve the 
resonances. In fact, the number of points in the tabulated cross
sections is so limited that even the radiative recombination rates 
cannot be computed with accuracy better than $\sim$5\%; (2) The He~I
cross sections do not contain all contributing series of resonances,
and the cross sections for other ions do not include any resonances 
at all; (3) the OP cross sections for He~I are total cross sections,
including photoionization into the n=2 states of the hydrogen target 
ion. However, under most conditions the bulk of the recombining ions
are in the ground state, so the calculation of recombination rates
should only use state specific photoionization cross sections to the ground state 
of the target ion.

To solve these problems with the OP He~I cross sections we remove the resonances 
by interpolation between points on opposite sides of the
resonances. Later, we truncate the cross sections at the energy
points right before
the first n=2 target threshold and extrapolate the cross sections
to higher energies using the quantum defect method of Peach (1967). 
In doing so, 
these quantum defect high energy cross sections are scaled to 
match the OP cross sections at the crossing point. 

For levels with n$>10$ and $l\le2$ we use cross sections from the
quantum defect method. For all levels with $l>2$ hydrogenic cross 
section from the method of Burgess and Seaton (1960a) were used. 

The state specific dielectronic recombination (DR) coefficients for n$\le10$ and $l\le4$
were taken from ADAS (Badnell \etal 1995). These ADAS rate coefficients are given in
tabulated form at temperatures between $(Z-1)^2\times 10^3$   and $(Z-1)^2\times 10^4$K which we fit with 
an expression of the form
\begin{equation}
\alpha_{DR}\times T^{3/2} =\sum_i \tau_i\exp{[(\epsilon_i-\epsilon_T)/kT]},
\end{equation}
for $i\le 4$. Here, $\epsilon_T$ is the ionization energy of the ion and
$\epsilon_i$ represents the energy of the autoionizing resonance $i$. The choice of such an expression and the values of $\tau_i$ and $\epsilon_i$ to fit the DR coefficients 
are explained is Section 4.1. The resulting fits to the ADAS data are 
accurate to better than 1\% in all cases. 

The DR coefficients for $n>10$ were estimated by n$^{-3}$ extrapolation 
of the coefficients for lower n. DR coefficients for states with angular
momentum, $l$, greater than 4 are expected to be very small and 
are neglected altogether. 

\subsection{Radiative transition probabilities}

The TOPbase data provides oscillator strengths, $f$, and transitions 
probabilities, $A$, for $n\le10$ and $l\le2$ for He-like ions. These 
transition probabilities are generally accurate to within a few per cent.
For He~I more accurate data was calculated by Kono and Hattori (1984).
Recently, Nahar and Pradhan (1999) computed fine structure $f$-values 
for Fe~XXV. Their results when summed over fine structure agree well 
with the TOPbase data.

For $n>10$ the A-values were computed using the Coulomb approximation
methods described in Smits (1991). 

Transition probabilities for the forbidden, $2~^3S-1~^1S$, and intercombination, $2~^3P-1~^1S$, 
transitions were taken from Mewe and Schrijver (1978a), which are based
on computations by Drake (1969, 1971a, 1971b), Johnson and Lin (1974),
and Drake and Dalgarno (1969).

\subsection{Collision strengths}

We adopt collisional data recommended by Dubau (1994) and Kato and    
Nakazaki (1989). These reviews consider data  for levels up to $n=5$
at most. Collisional rates for other $n$ changing transitions 
were calculated using the impact parameter method (Seaton 1962).
Rates for angular momentum changing transitions by collisions with 
electrons, protons, and alpha particles were computed using the
impact parameter method but with the modifications suggested by 
Hummer and Storey (1987).

\subsection{Collisional ionization and three body recombination}

Collisional ionization rates for excited levels are computed by using the semi-empirical 
formula by Sampson and Zhang (1988). This formula is based on extensive
Distorted Wave calculations of collisional excitation and is expected to 
be accurate for the He-like ions. However, the formula neglects the
contributions of collisional excitation-autoionization which,  in the case
of ionization from the ground state, 
sometimes 
dominate the ionization rate. Unfortunately, there is no data for 
excitation-ionization from excited levels. Thus we estimate that 
collisional excitation rates from excited states may have
uncertainties of around 50\%.
Three body recombination rates are obtained from the 
collisional ionization rates by means of detailed balance.
\begin{equation}
\alpha_{3B}={w_i\over w_{i+1}}{1\over2}\left({h^2\over2\pi mkT}\right)^{3/2}e^{I_{i}/kT}C_{i\infty}.
\end{equation}
were $C_{i\infty}$ is the collisional ionization rate coefficient, $w_i$ and $w_{i+1}$ are the statistical weights of the recombined
and recombining ions, $I_{i}$ is the ionization energy of the $i$
state.

\section{Calculation of Level Populations}

Under ionization-excitation equilibrium conditions the population of 
any level $n_i$ is given by 
\begin{eqnarray*}
n_i [\sum_{i\neq i}(A_{ij} + N_e C_{ij}^e + N_p C_{ij}^p) + 
+ N_e Q_{i\infty} ]& =
 \sum_{k< i} n_k A_{ki} + \sum_{l < i} n_l (N_e C_{li}^e +N_pC_{li}^p)\\
 & N^+N_e \alpha_i + N^+ N_e^2 C_{\infty i}
\end{eqnarray*}
where $A_{ij}$ is  the radiative transition rate from level 
$i$ to level $j$.
$N_e/N_p$ and $C_{ij}^e/C_{ij}^p$  are the density per unit volume
of electrons/protons and the transition rates by collisions
with electrons/protons respectively. $N^+$ is the density of the next
ionization stage of the element, i.e. the H-like ionic state in this case. $\alpha_i$ 
is the rate coefficient for 
recombination (radiative plus dielectronic) to level $i$. $C_{i\infty}$
and $C_{\infty i}$
are the collisional ionization and 3-body recombination rate 
coefficients respectively.
In principle, the sums in this equation includes energy levels 1
through infinity. However, in the present calculation it was found
sufficient to explicitply include levels up to $n=50$ only,
while cascades from higher levels were accounted for by 
extrapolation.

This recombination-collision-cascade problem was solved with  the
matrix
condensation technique (Burgess and Summers 1976) using the 
computer code developed by Smits (1991).

The population of all levels up to n=50 of every ion were calculated
for several temperatures covering the entire range between $10^3$ and
$10^9$K and 
nine electron densities from 10$^2$ to $10^{18}$ \cm3.

The high density calculations required special attention to account 
for the effects of lowering of the continuum and suppression of dielectronic
recombination.     
The treatment of these high density effects is described next. 

\subsection{High density lowering of the continuum}

The lowering of the continuum effect is very important under the high
plasma
densities considered here. This
effect comes about from cutting the high-$n$ orbitals of ions due to
particle packing, Debye shielding, Stark broadening, and collisional
broadening.
 Under the conditions of interest here particle
packing is the most important of these mechanisms. Particle packing
occurs when the mean
inter-nuclear separation in the plasma is smaller than the distance
from the nucleus to the high-$n$ ionic orbitals.
By comparing the mean inter-nuclear
separation in the plasma with the mean size of ionic orbitals of
principal quantum number $n$, one can define a continuum level as (Hahn
1997):
\begin{equation}
n_P = (1.8887\times 10^8 z/N^{1/3})^{1/2},
\end{equation}
where $z$ is the nuclear charge of the ion considered and $N$ is the
density of nuclei in the plasma in \cm3.

Debye shielding is the mechanism by which high-$n$ orbitals are shielded
from the electrostatic attraction to the nucleus by a high density of
free electrons.
According to this mechanism the continuum level is given by
(Hahn 1997):
\begin{equation}
n_D = 2.6\times 10^7 z^2 (T_e/N)^{1/4},
\end{equation}
where $T_e$ is in Kelvin and $z$ effective charge if the ion (i.e.
$z=Z-1$ for He-like ions).

Under a high concentration of singly charged ions in a plasma a
micro-electric-field is formed which will lead to Stark broadening
of the atomic levels. Then, for sufficiently high $n$-numbers
the atomic levels will merge with each other lowering the continuum. In
this case the continuum level is given by (Inglis \& Teller 1939):
\begin{equation}
n_S = [1.814\times 10^{26} z^6/N] ^{2/15}.
\end{equation}
For temperatures lower than $10^5$ K$z^2/n$ free electrons contribute to
the broadening through the static Stark effect. So the density $N$ in
the last
equation should include both positive and negative charges. At higher
temperatures the electrons contribute to the broadening by means of
collisions, but this is smaller than the Stark effect of the same
electrons at lower temperatures. Thus, only positive
charges are considered in this case.     

Furthermore, one may define $n_C$ as the minimum of $n_P, n_D,
$ and $n_S$.
Then, only orbitals with $n<n_C$ may be treated as bound, while
orbitals with $n>n_C$ are mixed with the continuum and must be
excluded from the calculation of the recombination process.
One can see that in the case of hydrogen like oxygen (z=8), for example, under conditions
of T=$10^5$ K and $N$=
$10^{16}$\cm3 only levels with $n\le 83$ are bound, at $N=10^{18}$
\cm3 only levels with  $n\le 38$ are bound, and at $N=10^{20}$ \cm3
only levels with  $n\le 18$ are bound.

\subsection{High density suppression of dielectronic recombination}

The effects of high electron densities and radiation fields on DR
were studied by Burgess and Summers (1969). 
DR occurs by means of three basic processes:          

(1) Resonance capture, autoionization: $X^{+z}(i)+e^-\leftrightarrow [X^{+(z-1)}(j,nl)]$;

(2) Stabilization: $[X^{+(z-1)}(j,nl)]\to X^{+(z-1)}(i,nl)+h\nu$; 

(3) Cascade: $X^{+(z-1)}(i,nl)\to X^{+(z-1)}(i,n'l')+h\nu$; 

\noindent{where} the brackets indicate an autoionizing state.

\noindent{Under} low electron density conditions the cascade process  
proceeds until the $X^{+(z-1)}$ ion reaches its ground state, so that
the recombination rate is just the total rate of all stabilizations.

High electron densities may affect the reactions (1), (2), or (3) 
as follows:

(a) The autoionizing states of the recombined ion, $[X^{+(z-1)}(j,nl)]$ 
in the right hand side of (1), could be affected by collisions with 
electrons. This would cause a redistribution among the different 
$l$-states, leading to an increase in the recombination rate coefficient.

(b) Collisional deexcitation could assist the stabilization (process
(2)) which would enhance the DR rate.

(c) Collisional ionization could reduce the number of bound states involved in stabilization process 
($X^{+(z-1)}(i,nl)$ in (2)). In other words, the lowering of the
continuum effect, discussed in the previous section, will 
suppress DR in similar fashion as it reduces the total radiative 
recombination rate.

For the range of densities considered here, $N_e\le10^{18}$ \cm3, the 
effects (a) and (b) can be neglected. This is because collisional 
transition rates, normally of the order of 
$N_e 10^{-5}/T_e^{1/2}$, have to compete with the much greater 
autoionization 
rates,
of order $10^{15}$ \s1. Furthermore, the net effect of high electron
densities considered here is to suppress DR. This has been calculated 
for some ions by various authors (e.g. Burgess and Summers (1969) and Summers 1972). On the other hand, Jordan (1969)
computed an empirical correction 
factors to DR as function of the maximum bound orbital of the ion before
the continuum. This approach allowed Jordan to calculate ionization balance
in dense plasmas 
in good agreement, $\sim5\%$, with the more elaborate computations of Summers
(1972). We adopt Jordan's method to compute DR suppression factors. 

\section{Ionization-excitation-spectra  models}            

The computed level populations of the He-like ions that were obtained
for  density vs. temperature grids allow us to construct
ionization-excitation-spectra (IES) models for each ion. These models are capable of 
simultaneously calculating the H/He-like ionization balance and the 
line emissivities under a variety of conditions. Our models are 
similar to those described for the He~I case by Cota (1987), Almog and
Netzer (1989) and Benjamin, Skillman, and Smits (1999). However, several
improvements have been made with respect to those models to allow
for accurate treatment of high densities, optical depth effects,
and convergence towards LTE.

Cota (1987) and Almog and Netzer (1989) explicitly consider a finite 
number of levels up to $n=n_{max}$ and adopt four fictitious levels 
that account for all more excited levels. The recombination cascades
from the fictitious levels to the explicit levels are accounted for 
by using averaged radiative transition probabilities 
weighted by the level population, which are assumed to be in LTE.
This assumption introduces 
errors in the calculated spectrum, particularly at low temperatures.
In the model of Benjamin \etal (1999) cascades from levels with     
$n> n_{max}$ to the lower are summed over to define "indirect"
recombination rates which are fit over a temperature range. This 
approach avoids the use of fictitious levels and yields higher 
accuracy under some conditions. However, the method of Benjamin \etal
introduces new problems when considering optical depth effects. 
Let us consider for example a model with indirect recombination rates
computed under optically thin conditions (Case A). When using this 
model to solve the Case B problem one would shut down all transitions
to n=1. Then, the total recombination obtained this way would be 
the total Case B recombination minus the fraction of recombination into
$n> n_{max}$ which cascades down to n=1. 
For even greater optical depths which affect transition to n=2,3, etc.
the error in the total recombination rate obtained by the Benjamin \etal
approach will become worse.

Our ionization-excitation model for He-like ions consists of all spectroscopic LS terms with n$\le$5 (i.e. 29 
levels), 2 so called ``superlevels", a level that represents the 
ground state of the H-like ion, and 6 doubly excited levels ($2s^2~^1S,
2s2p~^1P^o, ^3P^o,$ and $2p^2~^1D, ^1S, ^3P$). 

The superlevels are meant one for the singlets and one for the triplet
states and are built 
to account for all $5  < n\leq50$ levels. To do that we define: 

\noindent{-} recombination rates (RR+DR and three body recombination), $\alpha_s$:
$$\alpha_s=\sum_{m} \alpha_m,$$
- radiative transition probabilities, $A_{s\to i}$:
$$A_{s\to i}={\sum_{m} N_m A_{m\to i} \over \sum_{k} N_k},$$
- collisional deexcitation rates, $C_{s\to i}$:
$$C_{s\to i}={\sum_{m} N_m C_{m\to i} \over \sum_{k} N_k}.$$
Here, the sums include all levels with $5  < n\leq50$, $N_m$ represents the population of the level $m$ which is known
from the recombination calculation described in Section 3.
All of these rates for transitions involving the superlevels were 
computed for every pair of points in the temperature vs. density 
grid. From these values the transition rates 
at any temperature and density of interest can be obtained
with accuracy of approximately 2\% by linear interpolation. 
In addition, all the inverse processes to these rates are calculated by
means of detailed
balance. 

It is important to notice that by separating the collisional and 
radiative cascades from the superlevels one can  correctly treat the
problem of high density and high optical depth where radiative decays
from excited levels may be suppressed by self-absorption but cascades 
still occur via collisional deexcitation. 

The doubly excited levels in the IES models were included  to 
account for the most important satellite lines to H-like Ly$\alpha$
emission. Excitation energies, line wavelengths, radiative transition 
probabilities and autoionizing rates were taken from Vainshtein and 
Safronova (1978). We also included electron impact excitation rates
from bound to doubly excited levels and between doubly excited levels
from the calculations of Sampson, Goett, and Clark (1983) and Goett,
Sampson, and Clark (1983).

Explicit treatment of doubly excited levels introduces the 
additional practical complication that these levels will contribute 
to the DR to the bound levels. Thus, these DR contributions had to be
subtracted from the total DR to each bound level as taken from 
the ADAS database (Badnell \etal 1985). On the other hand, the detailed treatment of
doubly excited levels tends to improve the behavior of the models 
at high densities. This is because by considering collisional 
transitions involving these levels one accounts for the effects 
of redistribution of $l$ states and collisional assisted stabilization
in the DR process (see Section 3.2). In addition, collisional excitation
from bound to doubly excited levels results in excitation-autoionization
contributions to the collisional ionization rates which are missing 
in the rates of Sampson and Zhang (1988).

It is important, however, to be cautious about the 
predictions of satellite lines due to the large uncertainties in
some of the current atomic data, particularly  of 
the collisional rates for doubly excited levels 
which are based on Coulomb-Born-Exchange collision strengths that neglect
resonances. These resonances can dominate the collisional rates for 
doubly excited levels. For example, Bautista (2000) recently showed 
that in the case of doubly excited levels of Fe~XVI the resonances 
in the collision strengths would enhance the excitation rates by up
to three orders of magnitude. 
Furthermore, improved calculations of collision strengths and 
radiative and autoionization rates for He-like ions are currently underway.

\subsection{Convergence to LTE}

One important feature of the present IES models is their ability
to converge to LTE under appropriate condition. This characteristic 
is essential if they are to be applicable to conditions of dense 
plasmas and/or high radiation fields. 

For an IES model to properly converge to LTE every transition rate must 
be balanced by its detailed balance inverse. In the present models 
the transitions rates of interest and their inverse are:
\begin{center}
\begin{tabular}{rcl}
$collisional~deexcitation$&- &$collisional~excitation~cross~sections,$\cr
$spontaneous~transition~probabilities$&-&$induced~transition~probabilities,$\cr
$three-body~recombination$&-&$collisional~ionization~cross~sections,$\cr
$recombination~(RR+DR)$&-&$photoionization~cross~sections~including$\cr
 & & $\ autoionizing~resonances.$\cr
\end{tabular}
\end{center}
The collisional and radiative transition rates are readily
balanced from the effective collision strengths, $\Upsilon$, and
A-values. Three-body recombination and collisional ionization balance 
each other through Eqn.(2).
The balance between recombination (RR+DR) and photoionization is
somewhat more complicated to achieve due to the fact we have only 
background photoionization cross sections (i.e. without autoionizing 
resonances). These cross sections balance only the RR component of
the total recombination. To balance DR one needs to work backwards 
from the DR rates to find at least the dominant autoionizing resonances.

Let us consider a photoionization cross section with zero background 
and narrow  autoionizing resonances which can be approximated by 
delta functions at photon energies $\epsilon_i$, i.e.
\begin{equation}
\sigma(\epsilon) =\sum_i a_i\delta(\epsilon-\epsilon_i).
\end{equation}
From this cross section and using the Milne relation the recombination rate is          
\begin{equation}
\alpha={w_z\over w_{z-1}}\sqrt{2\over \pi} {h\over c^2}{1\over
(m_ekT)^{3/2}}\sum_i a_i \epsilon_i^2\exp{[-(\epsilon_i-\epsilon_T)/kT]},
\end{equation} 
where $w_z$ and $w_{z-1}$ are the statistical weights of the recombined
and recombining levels respectively and $\epsilon_T$ is the ionization threshold energy.
This equation is  equivalent to Equation (1), used to 
fit the ADAS DR rates,  if one defines 
\begin{equation}
\tau_i\equiv {w_z\over w_{z-1}} 1.167\times 10^{-10} \epsilon_i^2
a_i,
\end{equation}
for $\epsilon_i$ in Ryd and $a_i$ in Mb.
The choice of the location of the autoionizing resonances comes from 
the realization that the most prominent resonances are those that belong 
to the Rydberg series converging to the n=2 states of the H-like recombining
ions. These thresholds are located at energies of  $Z^2(1-1/4)$ Ryd 
above $\epsilon_T$. Thus, the resonances are approximately $(Z-1)^2/\nu$ Ryd
below the threshold, with $\nu$=2, 3, 4, 5, ... Thus,
\begin{equation}
\epsilon_i-\epsilon_T={3\over4}Z^2-{(Z-1)^2\over (i+1)}.
\end{equation}
In practice, we found that at most four
resonances are needed to fit the ADAS DR rates to an accuracy better 
than 1\%. Then, these resonances were added to the background photoionizations
cross section in order to guarantee that the models converge to LTE 
under proper conditions.

\section{Results}

The present section reports the results of the present IES models. We
start by comparing our effective recombination coefficients with other 
calculations. 
Then we focus on the case of coronal ionization 
equilibrium and discuss various line ratio diagnostics for optically  
thin plasmas. 
Finally, the spectra from photoionized plasmas is discussed.

\subsection{Recombination to n=2 terms}

Figs. 1(a)-(b) show the effective recombination rate coefficients 
(direct recombination and cascades) vs.
temperature for the n=2 terms $2~^3S,$ and $2~^3P^o$
respectively. The rates are shown for various He-like ions (O~VII, S~XV,
and Fe~XXV). 
Also in the same figures we show the rate coefficients from Mewe and 
Schrijver (1978a).

One can see from these figures that the present recombination rates 
differ significantly from those of Mewe and Schrijver. Particularly
problematic is the high temperature where recombination is dominated 
by DR. Here the discrepancies between Mewe and Schrijver and the 
present calculation can exceed a factor of four.

Also problematic are the recombination coefficients of Mewe and
Schrijver for the $2~^1S$ state which are systematically overestimated 
by a factor of $\sim3$. This is shown in Fig. 2 (upper-left  panel).
Confirmation about the Mewe and Schrijver coefficients being overestimated can be obtained from the He~I case.
At T=$10^4$K the direct recombination rate coefficient to $2~^1S$
is $5.55\times 10^{-15}$ cm$^{-3}$s$^{-1}$ (see Osterbrock 1989 and
Benjamin \etal 1999)
and our present total rate coefficient, including cascades, to this state
is $6.55\times 10^{-15}$ cm$^{-3}$s$^{-1}$, which is 16\% lower than
sum of ``direct" and  ``indirect" recombination rates of Benjamin \etal. 
On the other hand,
the Mewe and Schrijver is $1.99\times10^{-14}$ cm$^{-3}$s$^{-1}$ which
is clearly too high by a factor of 3.

Fig. 2 presents the effective recombination coefficients for n=2
terms against ionic nuclear charge ($Z$) for temperatures of
$10^4\times(Z-1)^2$ K and low electron density. 
Also in this figure, are the results of Mewe and
Schrijver (solid curves) and the  hydrogenic extrapolation with $(Z-1)$ from neutral helium (dashed curves), i.e.
$$\alpha(T,Z)=(Z-1)\alpha(T/(Z-1)^2,1).$$

There is good agreement between the present
results and those of Mewe and Schrijver for $2~^1P$ and $2~^3S$, but
some differences are present in for the $2~^3P$ term as well as 
the systematic factor of 3 discrepancy for the $2~^1S$ term.
On the other hand, the
hydrogenic extrapolation of the rate coefficients reproduces very well
the recombination rates for $2~^1S$ and $2~^3S$ but it systematically
overestimates the rate coefficients for the $2~^1P$ and $2~^3P$
terms. For the latter term, the hydrogenic extrapolation and the results
of
Mewe and Schrijver are almost indistinguishable. 

It is important to note that the $2~^3S$ level is responsible for the 
well known forbidden ($f$) line of He-like ions which is used in 
density and temperature line ratio diagnostics (see Section 5.2).
Furthermore, the overestimation of the effective recombination rate 
to this level by MS yields erroneous diagnostics. 

\subsection{Density and Temperature Diagnostics}

The emission lines that originate from transitions from n=2 levels to
the ground state are the most intense in the spectra of He-like ions.
These are three resolvable features in medium resolution spectra, i.e.
the forbidden ($f$) $2~^3S-1~^1S$, intercombination ($i$) 
$2~^3P-1~^1S$, and resonant ($r$) $2~^1P-1~^1S$ transitions. Strictly
speaking, the $i$ feature is composed of two fine structure lines, i.e.
$2~^3P_2-1~^1S_0$ and $2~^3P_1-1~^1S_0$. Historically, these lines are 
also referred to as $x$ and $y$ lines, and the $f$ and $r$ lines are also
called $z$ and $w$.

Under collisional ionization conditions line ratios among the $f,\ i$,
and $r$ features are useful density and temperature diagnostics.
In photoionized plasmas, however, these diagnostics are misleading
(Section 5.3).    
The intensity ratio $R={I_f/I_i}$ is sensitive to 
electron density above a critical value. This is because while the
$2~^3S$ and $2^3P$ states are populated at nearly constant proportions
for a wide rage of conditions the population of $2~^3S$ can be excited
to the $2~^3P$ state for densities such $C(2~^3S-2~^3P)N_e \ge
A(2~^3S-2~^1S)$. The ratio $G=(I_f+I_i)/I_r$
is particularly sensitive to temperature. This is because at high 
temperatures all the 
lines are mostly collisionally excited, but at lower 
temperatures recombination dominates the population of the triplet
states. 

Blends of satellite lines with the $f$ and $i$ lines complicate 
the picture as they too have a temperature dependence. The satellite
lines come from transitions of the type $1s^2nl-1s2snl$. Dielectronic 
recombination is the most important excitation mechanism for the
majority of these lines, but for some satellites collisional inner-shell
excitation is important. The intensity ratio of dielectronic satellites
to the the $r$ line is roughly proportional to $T^{-1}$. 

Fig. 3 shows the $R$ ratios vs. \ne for several ions. The density sensitivity 
of the $R$ ratios starts at around $10^{9}$ \cm3 for O~VII 
and around $10^{16}$ \cm3 for Fe~XXV. 
The systematic behavior of the $R$ and $G$ ratios along the helium
isoelectronic sequence were explored by Pradhan (1982).
Under low density conditions the $R$ ratios exhibit a relatively weak 
temperature dependence. 

$G$ ratios vs. temperature plots are shown in Fig. 4 (solid curves). These ratios
present strong temperature dependence which makes them good temperature
diagnostics. 
Also in this plots we show the ratios when the satellite contributions
are included (dashed curves). These satellite lines lie to the blue side but very close
to the $f$ and $i$ lines. Thus satellites will enhance the apparent G
ratio as measured in low to medium resolution spectra. As a consequence,
neglecting the contributions of unresolved satellites in the spectra
may result in underestimation of the temperature by up to one order of
magnitude.

\subsection{Emission from Photoionized Plasmas}

The traditional line ratio diagnostics discussed above are only
applicable to coronal plasmas. On the other hand, photoionization 
may be the dominant ionization mechanism in a  variety of astronomical 
plasmas. The nature of the ionization mechanism (photoionization or
collisional ionization) of the observed plasma 
must be established before diagnostics can be performed.   
Fig. 4 shows the G ratio vs. temperature under photoionization 
conditions (dotted curves). 
It can be seen that observed ratio from photoionized plasmas departs 
considerably from that in the coronal case indicated by the solid curve. Thus, failure to recognize
photoionization contributions to the emitted spectra will result in
misleading diagnostics.

Direct evidence of photoionized plasmas can be found in the strength 
of lines from the $n=3$ levels. The strongest of these lines are:
$3~^1P-1~^1S$ and $3~^3D-2~^3P$. Table 1 compares the relative strengths 
of these lines to the $r$ line in a coronal and photoionized
plasmas. The results are given for a $2\times10^6$K coronal plasma,
near the temperature of maximum O~VII abundance, and at a
temperature of $10^5$K, near the thermal equilibrium temperature,
for a photoionized plasma. 
In general, lines from $n=3$ levels are very weak or undetectable
in coronal plasmas, while photoionized plasmas do produce these lines
with strengths of up to $\sim25\%$ of the $r$ line.   

\begin{deluxetable}{cll}
\small
\footnotesize
\scriptsize
\tablewidth{36pc}
\tablecaption{Line intensities relative to the $r$ line for coronal and
photoionized O~VII}
\tablehead{
\colhead{Transition} &
\colhead{Coronal (T=$2\times10^6$K)} &
\colhead{Photoionized (T=$10^5$K)} }
\startdata
$2~^1P-1~^1S$ & 1.0   & 1.0 \nl
$2~^3S-1~^1S$ & 0.57  & 3.47 \nl
$2~^3P-1~^1S$ & 0.16  & 0.98 \nl
$3~^3P-2~^3S$ & 0.002 & 0.10 \nl
$3~^3D-2~^3P$ & 0.001 & 0.27 \nl
$3~^1P-1~^1S$ & 0.014 & 0.20 \nl
\enddata
\end{deluxetable}

Under photoionization conditions a good temperature diagnostic can be
obtained from the ratio $I(3~^1P-1~^1S)/I(3~^3D-2~^3P)$ (P ratio
hereafter). This ratio is plotted against temperature in Fig. 5.
This ratio conforms a  reliable diagnostic without blending with 
satellite lines and with very high critical densities for collisional
deexcitation.

\section{Conclusions}

We have carried out detailed calculation of recombination spectra of
the He-like ions He~I, C~V, N~VI, O~VII, Ne~IX, Mg~XI, Si~XIII, S~XV,
Ar~XVII, Ca~XIX, and Fe~XXV. The calculations include radiative and 
dielectronic recombination, collisional ionization and three-body
recombination, and the effects of high density lowering of the
continuum. 

These calculations allow us to build
ionization-excitation-spectral models for fast computation of the
spectra for modeling and/or spectral diagnostics. 
This models include all levels up to $n=5$ and the most important
satellite lines to the H-like spectra.
The IES models are applicable to densities of up to $10^{18}$ \cm3
and converge to LTE under proper conditions.

The results of the IES models are compared with calculations by other
authors. It is found the present effective recombination rates differ
significantly from earlier estimates by Mewe and Schrijver (1978a).
For example, we find that the Mewe and Schrijver recombination rates 
for the $2~^1S$ state are overestimated by a factor of three.
The $2~^1S$ level decays to the ground state either two-photon 
continuum or collisional transitions. Thus, the discrepancy found in
the recombination rates for this level will be important in trying
to simultaneously model line and continuum emission. The 
difference in recombination to the $2~^1S$ level may also affect 
the line emission spectrum under high density or high optical depth
conditions where the $2~^1S$ level contributes to the population of
other levels.

Further, we present several line ratio diagnostics under coronal and
photoionization conditions.
It is shown that blends of satellite lines with the traditionally studied forbidden and 
intercombination lines can considerably enhance the apparent $G$ ratio
in coronal plasmas. 
Also, we find that the traditional $R$ and $G$ line ratio diagnostics used for coronal 
plasmas are misleading when the plasma is photoionized. Such
photoionized plasmas can be recognized by the presence of lines arising
from $n=$3 levels in the spectra. Moreover, these lines can used for 
reliable temperature diagnostics. 

The present IES models, data files and a computer code, is available upon
request to the authors. These models can be implemented in spectra
modeling codes or used as a stand alone tool for spectral diagnostics.
In this sense, the present IES models for He-like ions and similar
ones for H-like (Bautista \etal 1998) are already implemented in the 
photoionization modeling code XSTAR v.2 (Kallman and Bautista 2000).
For isoelectronic sequences other than H and He-like
XSTAR v.2 accounts only for direct recombination to every level, but
calculations of 
more detailed recombination models like those described in this paper
are underway.

\vspace{0.8in}

\section*{Acknowledgments}

This work has been supported in part by a grant from the Smithsonian 
Observatory for the Emission Line Project.

\clearpage 

\begin{figure}
\psfig{file=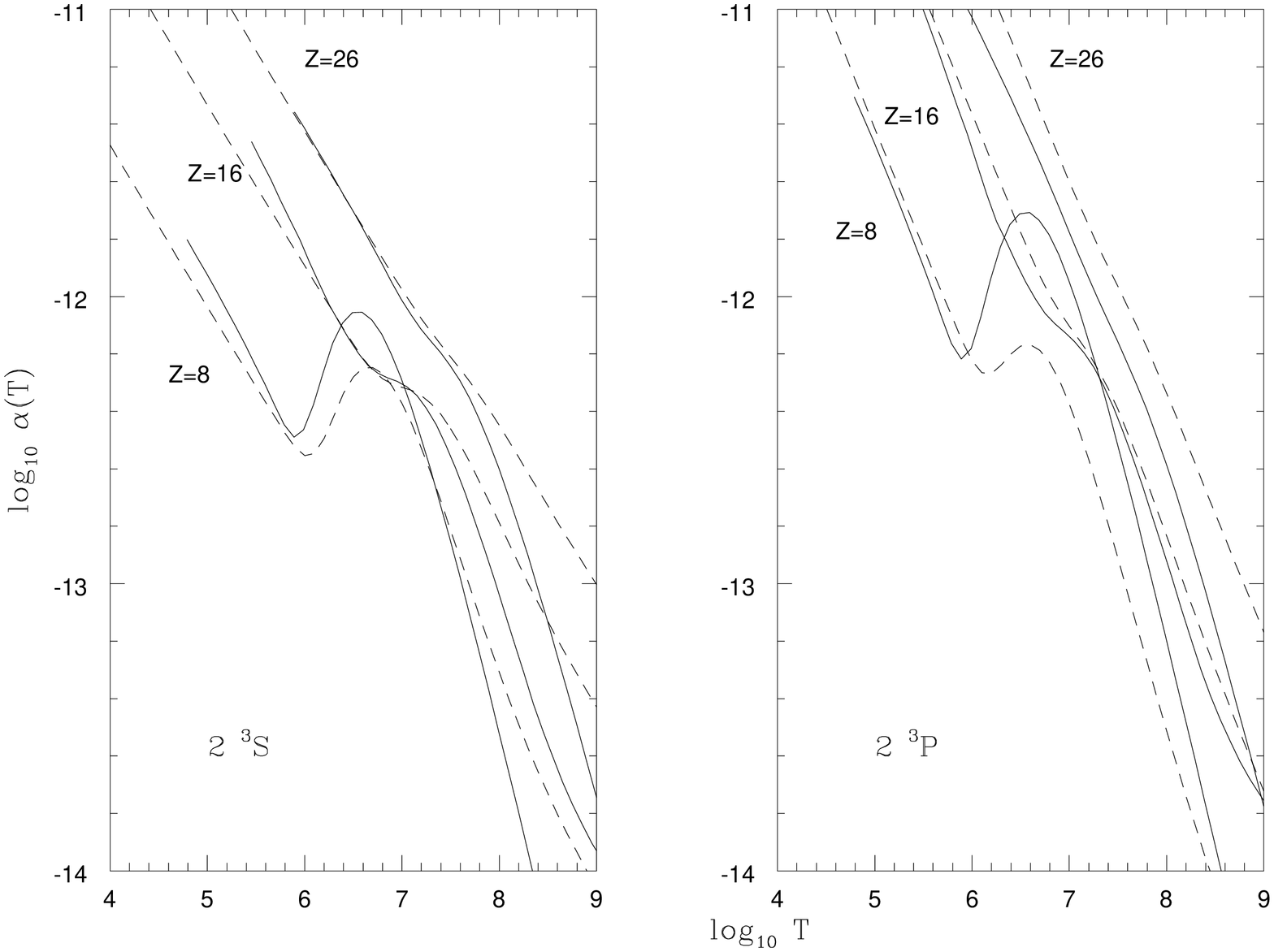,width=6in,angle=-90}
\caption{Comparison of effective recombination rate coefficients (direct
recombination and cascades) against temperature for $n=2$ term between
present 
calculations
(solid lines) and estimates by Mewe and Schrijver (1978; dashes lines).}
\end{figure}

\clearpage
\begin{figure}[h]
\psfig{file=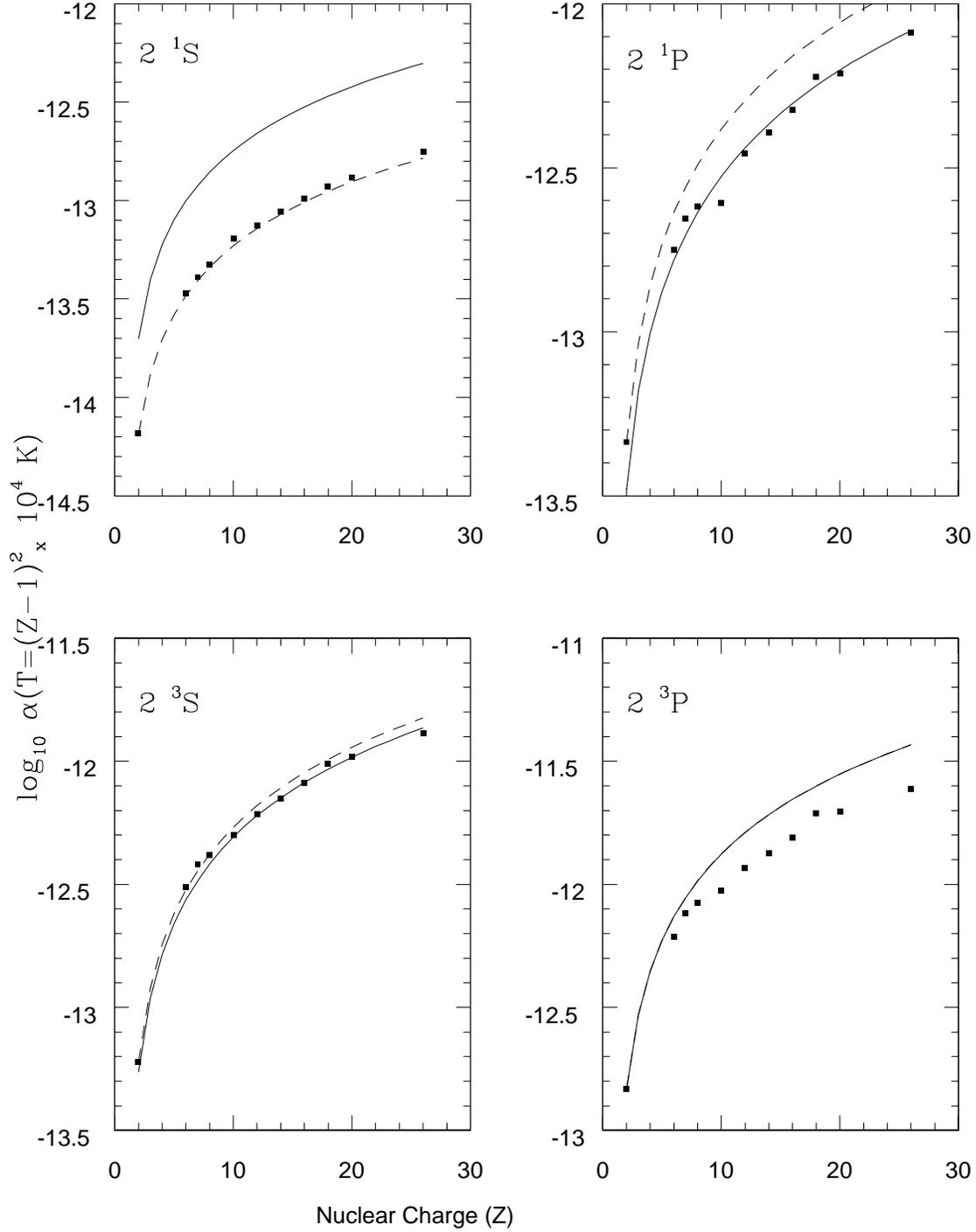,width=6in}
\caption{Effective recombination coefficients at T=$10^4\times$(Z-1)$^2$
K vs.
nuclear charge Z. The square points represent the present results,
the solid lines represent the estimates of Mewe and Schrijver (1978),
and the dashed lines are hydrogenic extrapolations of rate coefficients
from He~I
 (see text).} 
\end{figure}

\clearpage
\begin{figure}[h]
\psfig{file=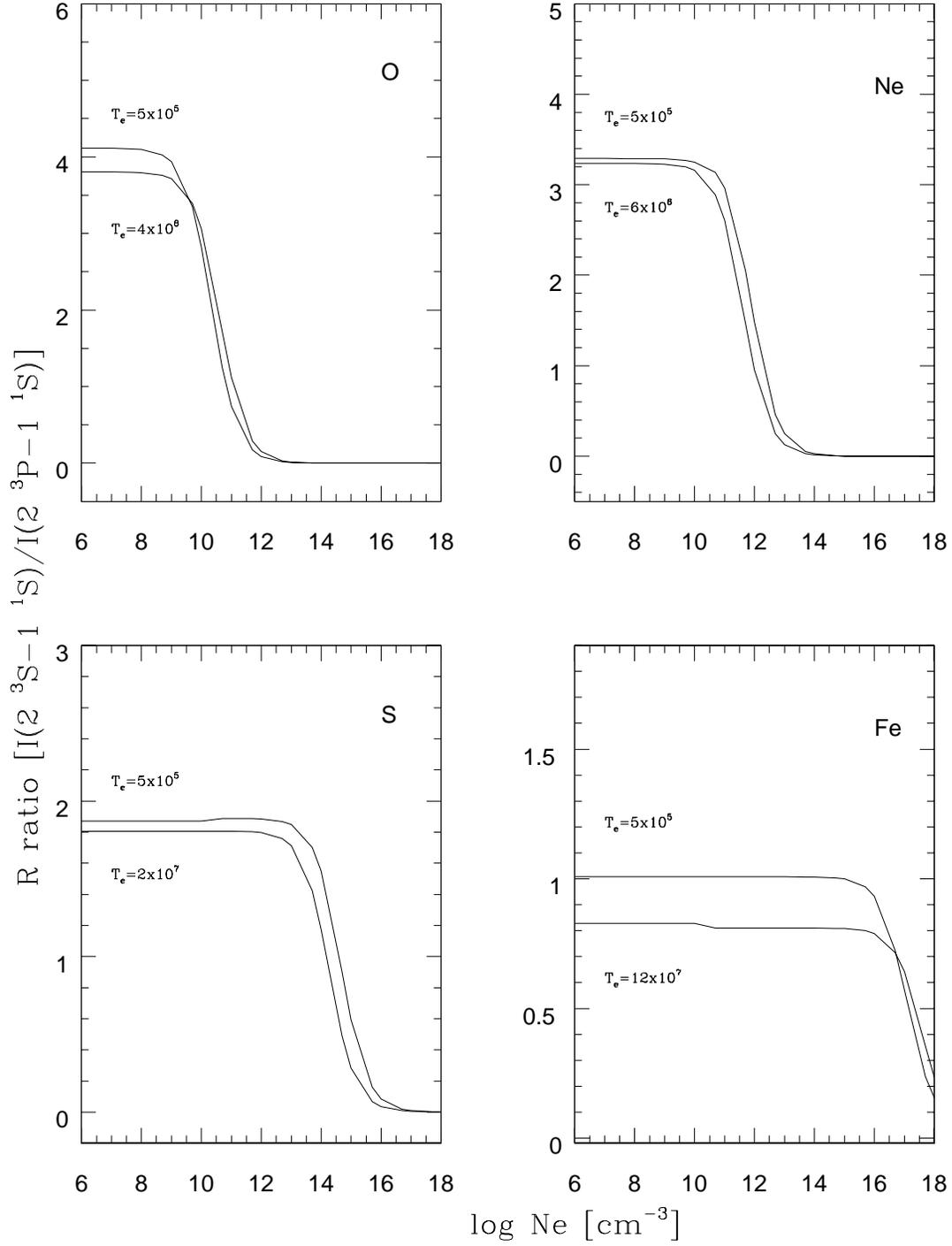,width=6in}
\caption{The R ratios (forbidden to intercombination line intensity
ratios)
vs. electron density.}
\end{figure}

\clearpage
\begin{figure}[hp]
\psfig{file=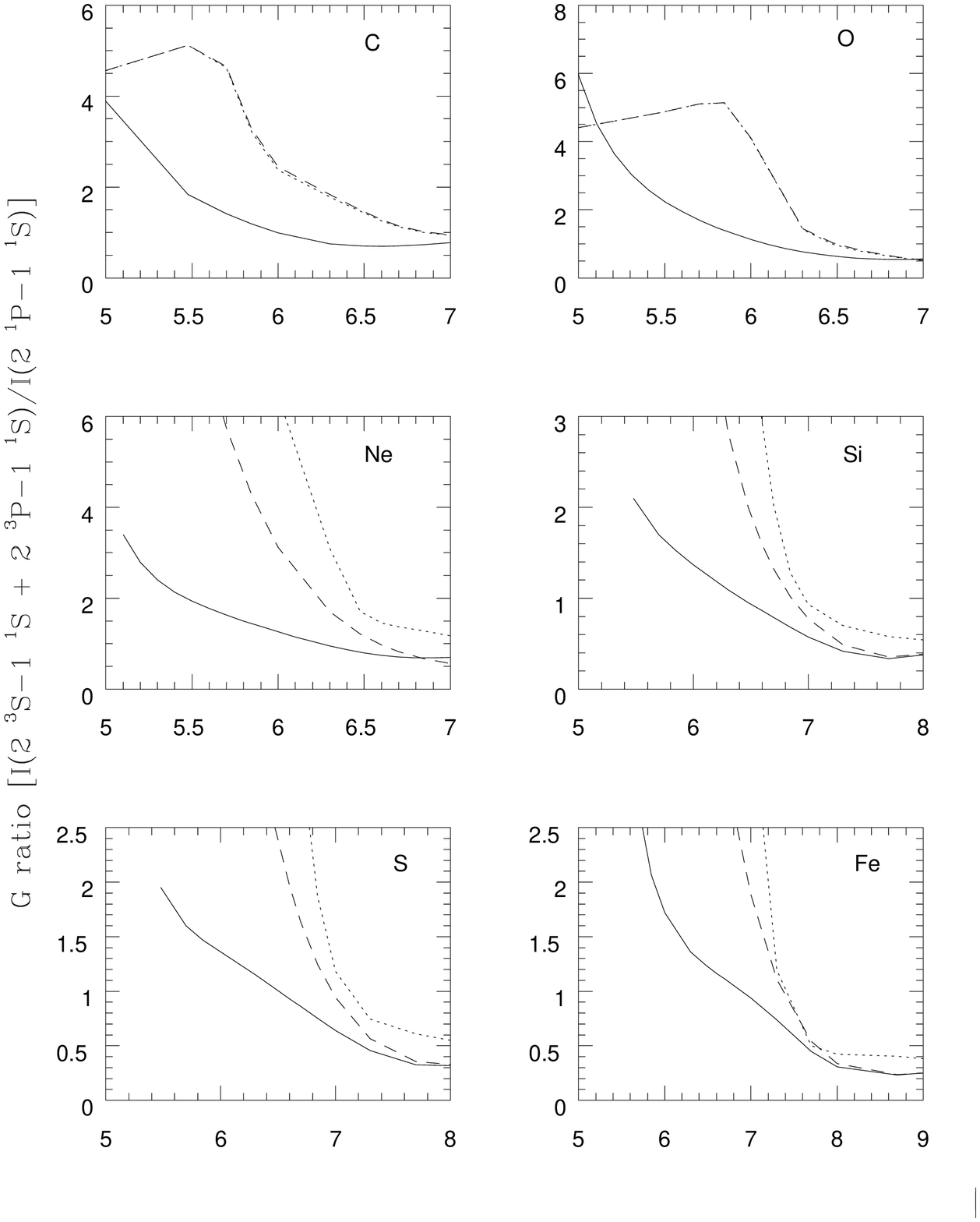,width=6in}
\caption{The G ratios (forbidden plus intercombination to resonant line
intensity ratios) vs. temperature. The solid curves represent the
the ratios under coronal ionization conditions. The dashed lines show
the ratios
 including satellites line contributions. The dotted lines represent the
G ratios
 under photoionization
conditions.} 
\end{figure}

\clearpage
\begin{figure}[hp]
\psfig{file=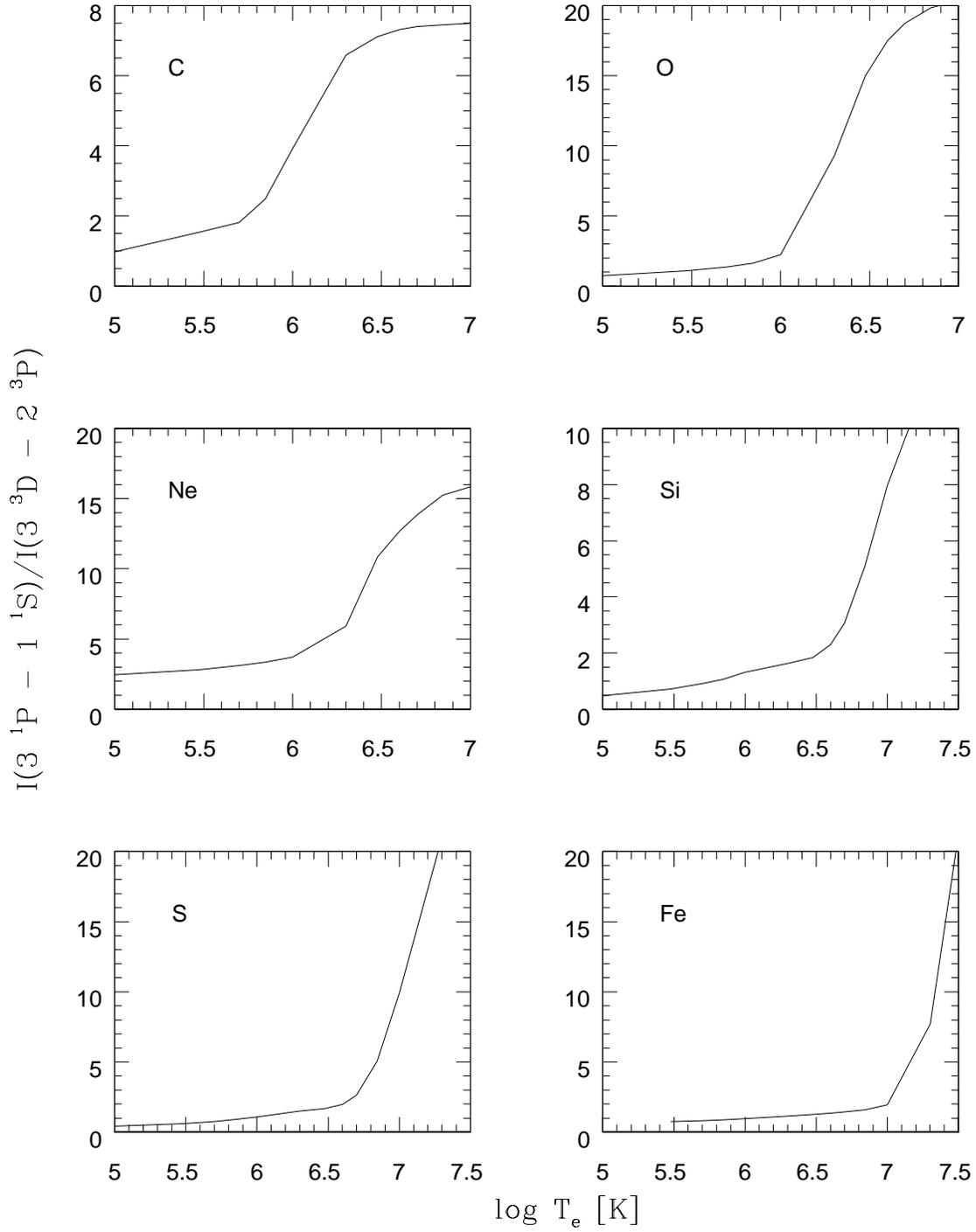,width=6in}
\caption{The P ratios ($I(3~^1P-1~^1S)/I(3~^3D-2~^3P)$) vs.
temperature.}
\end{figure}

\end{document}